\documentclass[pra,twocolumn,amsmath,amssymb,superscriptaddress,showpacs]{revtex4}
\usepackage{graphics}
\usepackage{epsfig}
\usepackage{bbm}
\usepackage[latin1]{inputen c} % Umlaute!

\newcommand{\be}{\begin{equation}}
\newcommand{\ee}{\end{equation}}
\newcommand{\eea}{\end{eqnarray}}

\newcommand{\exs}[1]{\ensuremath{\langle{#1}\rangle}}

\newcommand{\eins}{\openone}

\newcommand{\ketbra}[1]{\ensuremath{| #1 \rangle \langle #1 |}}

\newcommand{\ket}[1]{\ensuremath{|#1\rangle}}

\newcommand{\kommentar}[1]{}
\newcommand{\trace}{{\rm Tr}}

\begin{document}
\title{Genuine three-partite entangled states with a local hidden variable model}
\date{\today}
\begin{abstract}
We present a family of three-qubit quantum states with a basic
local hidden variable model. Any von Neumann measurement can be
described by a local model for these states. We show that some of
these states are genuine three-partite entangled and also
distillable. The generalization for larger dimensions or higher
number of parties is also discussed. As a byproduct, we present
symmetric extensions of two-qubit Werner states.

\end{abstract}

\author{G\'eza T\'oth}
\email{toth@alumni.nd.edu} \affiliation{Research Institute for Solid
State Physics and Optics, Hungarian Academy of Sciences,  P.O. Box
49, H-1525 Budapest, Hungary} \affiliation{ Max Planck Institute for
Quantum Optics, Hans-Kopfermann-Stra{\ss}e 1, D-85748 Garching,
Germany}
\author{Antonio Ac\'{\i}n}
\affiliation{ICFO-Institut de Ci\`encies Fot\`oniques,
Mediterranean Technology Park, 08860 Castelldefels (Barcelona),
Spain}


\maketitle

% 03.65.Ud                  Entanglement and quantum nonlocality
% 03.67.Mn                  Entanglement in Quantum Information
% 03.67.-a                  Quantum information

%%%%%%%%%%%%%%%%%%%%%%%%%%%%%%%%%%%%%%%%%%%%%%%%%%%%%%%%%%%%%%%%%%%%%%

\section{Introduction}
One of the most striking characteristics of quantum mechanics is
nonlocality. If quantum mechanics could be described by a local
hidden variable model (LHV) then
the values measured for multi-particle correlations could be
reproduced assuming that all measurable single-particle operators
had already a value before the measurement. Bell showed that there
are quantum states for which the many-body correlations cannot be
explained based on this assumption \cite{B64}. However, such
correlations arise only for {\it some} entangled quantum states
while for separable states the correlations can always be mimicked
by a LHV model \cite{W89}.

Proving that the measurement results on a quantum state cannot be
obtained from a LHV model is done by finding a Bell inequality
which is violated by the state \cite{B64}. However, this is
difficult since the determination of all Bell inequalities is a
computationally hard problem \cite{Pitowski}. To prove that any
measurement on a given state can be described by a LHV model is
perhaps even more challenging. This is because in order to do that
one has to find a LHV model for \emph{any number of
arbitrary operators} measured at each party.

Due to the difficulty of the problem, LHV models have a quite
limited literature. The first and most fundamental result of the
subject was presented by Werner in Ref. \cite{W89}. He described a
LHV model for arbitrary von Neumann measurements for some
$U\otimes U$ symmetric bipartite states \cite{W89}. For the qubit
case these are of the form
\begin{equation}\label{wstates}
    \rho_W=p\,\ketbra{\psi^-}+(1-p)\frac{\eins}{4},
\end{equation}
where $\ket{\psi^-}=(\ket{01}-\ket{10})/\sqrt 2$ is the singlet
state. In the same paper Werner also gave the first modern
definition of quantum entanglement, thus distinguishing it from
nonlocality. Indeed, states $\rho_W$ are entangled for $p>1/3$ and
local for $p\leq 1/2$~\cite{W89}. It is hard to overestimate the
importance of these results for the development of quantum
information science. Later, Barrett obtained a model for general
measurements, also called positive operator valued measures (POVMs),
for a subset of Werner states~\cite{B02}. LHV models were also
constructed for finite number of settings for states with positive
partial transpose exploiting symmetric extensions \cite{TD02}. Apart
from their fundamental interest, these results are also relevant
from a quantum information theory viewpoint. Simulating entanglement
by classical means (e.g., Ref.~\cite{MB01}) sheds light on the power
of entanglement as information resource. In this context, those
quantum states for which the correlations can be reproduced by a LHV
model are useless for communication tasks, since they do not provide
any advantage over shared classical randomness~\cite{BZ04}.

New and interesting open questions on the relationship between
nonlocality and entanglement appear in the multipartite scenario.
Recall that multipartite entanglement is known to be inequivalent to
bipartite entanglement \cite{BPRST}. Moreover, genuine multipartite
entanglement is the property most often detected in experiments
(e.g., Ref.~\cite{Exp}). We know that the Bell inequality violation
required for genuine multipartite entanglement (i.e., when all
parties are entangled with each other \cite{BISEP}) increases
exponentially with the number of parties \cite{Nagata}. Hence one
could expect that entanglement of this type for large enough number
of parties provides a sufficient condition for a state to be
nonlocal. However, beyond the bipartite case, the connection between
nonlocality and entanglement remains largely unexplored. Indeed, LHV
models for {\it multipartite entangled systems} are still missing.

In this paper, we present a one-parameter family of three-qubit
states whose correlations for von Neumann measurements can be
reproduced by a LHV model. Thus these states do not violate any Bell
inequality. Then, we prove that, remarkably, some of these states
have genuine three-qubit entanglement \cite{BISEP} and we also show
that they are distillable. To our knowledge, these are the first
examples of genuine multipartite entangled states allowing for a
local description. The generalization of the construction to other
situations, more parties or higher dimensional systems, is also
discussed.

Before proceeding, let us introduce the notation. We denote von
Neumann measurements  on $n$ parties $A,$ $B,$ $C,$ $...$ as
$M_{A},$ $M_{B},$ $M_{C},$ $...$. The spectral decomposition of
$M_A$ is given as $M_A=\sum_{k=1}^{d} \alpha_k P_k.$ In the case
of qubits, that we mostly consider in this work, $\alpha_1=+1$ and
$\alpha_2=-1$, while $M_A=\hat n_A\cdot\vec\sigma$, where $\hat
n_A$ is the normalized vector defining the direction of the von
Neumann measurement and $\vec\sigma$ is the vector of Pauli
matrices, $\vec{\sigma}=(\sigma_x,\sigma_y,\sigma_z)$.

\section{Two-qubit case}
The key point for the construction of our LHV model for three
qubits is an alternative derivation of Werner's original result
for two qubits. This new derivation has the advantage of being
easily generalizable to the case of three qubits. Consider the
two-qubit operators
\begin{eqnarray}
\rho^{(2,c)}:=\int_{\omega \in \mathbb{C}^2,|\omega|=1 }
M(d\omega) \varrho_\omega \otimes \rho_\omega , \label{decomp}
\end{eqnarray}
where
\begin{eqnarray}
\varrho_\omega&:=&\frac{1}{2}
[\eins-c\sum_{k=x,y,z}{\rm sign}(\exs{\sigma_k}_\omega)\sigma_k],\nonumber\\
\rho_\omega&:=&\ketbra{\omega}. \label{rhoomega}
\end{eqnarray}
Here $\ket{\omega}$ is a two-element state vector and $M$ is
% a $\sigma$-additive normalized measure.
the unique probability measure invariant under all single-qubit
unitary rotations. Direct calculation shows that $\rho^{(2,c)}$ are
Werner states~(\ref{wstates}) with $p=c/2$.
Based on this construction, the following statement can be made\\
{\it Theorem 1~\cite{W89}: There exists a LHV model for von
Neumann measurements on states $\rho^{(2,c)}$ for $c\leq 1$.}

\begin{figure}
%\narrowtext
\centerline{\epsfxsize 3in \epsffile{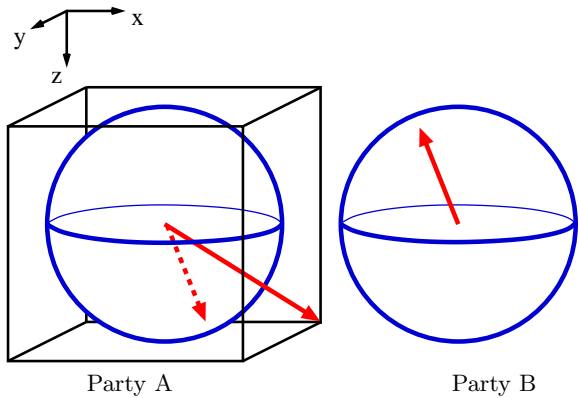} }
\centerline{ \hskip 0.3cm Party A \hskip 3.6cm Party B}
\caption{Schematic representation of our two-qubit hidden variable
model. Party B receives a standard Bloch vector. The dashed arrow
on the left hand side points opposite to this vector. Party A
receives a vector pointing to one of the eight vertices of a cube
tangent to the Bloch sphere. The vertex is chosen such that the
overlap with the dashed vector is maximal.} \label{fig_spheres}
\end{figure}

Before starting the proof, let us explain the intuition behind it.
First of all, note that one can restrict the analysis to $c=1$. In
this case, the decomposition (\ref{decomp}) can be understood as a
sort of local model for which party $B$ receives a standard Bloch
vector, $\hat n_\omega$, while $A$ receives one of the vectors
pointing to the vertices of a cube. This cube is tangent to the
Bloch sphere for the Bloch vectors which point to the directions
$\pm x$, $\pm y$ and $\pm z$ (see Fig.~\ref{fig_spheres}). Actually,
$A$ receives the vector with maximum overlap with $-\hat n_\omega$.
Using the standard trace rule $\exs{M_A \otimes M_B}=\int M(d\omega)
Tr(M_A\varrho_\omega)Tr(M_B\rho_\omega)$, we have a LHV model for
$M_A=\pm \sigma_{x/y/z}$ and arbitrary $M_B.$ The choice of $M_A$ is
restricted since if party $A$ chooses other operators to measure,
for some $\omega$ she would obtain $|\exs{M_A}_{\varrho_\omega}|>1$.
We then say that party $B$ has a physical qubit, while $A$ is
receiving a non-physical Bloch vector. As we have already said,
state $\rho^{(2,1)}$ is a Werner state, i.e., it is invariant under
transformations of the form $U\otimes U$ where $U$ is an arbitrary
unitary matrix. Using this symmetry, we can construct a LHV model
for all measurements. The detailed proof goes as follows:

{\it Proof of Theorem 1.} The goal is to find a LHV model for the
state (\ref{decomp}) with $c=1$, that is to write its correlations
as
\begin{equation}
\trace(M_A\otimes M_B\,\rho^{(2,1)})=\int_{\omega} M(d\omega)
\exs{M_A}_{\omega} \exs{M_B}_{\omega}, \label{corr}
\end{equation}
where $\exs{M_{A/B}}_{\omega}$ are the expectation values of
$M_{A/B}$ if the value of the hidden variable is $\omega$, and we
require $|\exs{M_A}_{\omega}|,|\exs{M_B}_{\omega}|\leq 1$.
Identifying the sub-ensemble index $\omega$ in Eq.~(\ref{decomp})
with the hidden variable in Eq.~(\ref{corr}),
%and using the standard trace rule
one has a LHV model with
\begin{eqnarray}
\exs{M_A}_\omega&=&\trace(M_A \varrho_\omega)=-\frac{1}{2}
\trace[M_A\sum_{k=x,y,z}{\rm sign}(\exs{\sigma_k}_\omega)\sigma_k],\nonumber\\
\exs{M_B}_\omega&=&\trace(M_B \rho_\omega)=\trace(M_B
\ketbra{\omega}). \label{lhv1}
\end{eqnarray}
It is clear that this model
works only if $M_A=\pm \sigma_{x/y/z}$.

Now we modify our LHV model in order to allow arbitrary operators
$M_A$ of the type $\hat{n}_A\vec{\sigma}.$ Such an operator can be
written in the form
\begin{eqnarray}
M_A=U_A^\dagger \sigma_z U_A.
\end{eqnarray}
We can take advantage of the invariance of Werner states under
transformations of the form $U\otimes U$, so $\exs{M_A \otimes
M_B}=\exs{\sigma_z^{(A)}\otimes M_B'}$, where
$M_B'=U_AM_BU_A^\dagger.$ Hence
\begin{eqnarray}
\exs{M_A}_\omega&=&\trace(\sigma_z \varrho_\omega)
=-{\rm sign}[\trace(\sigma_z\ketbra{\omega})],\nonumber\\
\exs{M_B}_\omega&=&\trace(M_B'
\rho_\omega)=\trace(U_AM_BU_A^\dagger \ketbra{\omega}).
\label{lhv2}
\end{eqnarray}
Indeed, substituting Eq.~(\ref{lhv2}) into Eq.~(\ref{corr})
reproduces all two-qubit correlations $\exs{M_A\otimes M_B}.$
However, Eq.~(\ref{lhv2}) is not a LHV model for arbitrary $M_A$ and
$M_B$ yet. This is because $\exs{M_{B}}_\omega$ depends on $U_A$,
i.e, it depends on what operators are measured on party $A$. This
dependence can be removed by defining $\ket{\omega'}=U_A^\dagger
\ket{\omega}.$ We obtain the desired LHV model with $\omega'$ as the
hidden variable given as
\begin{eqnarray}
\exs{M_A}_{\omega'}&=&-{\rm sign}[\trace(M_A\ketbra{\omega'})],\nonumber\\
\exs{M_B}_{\omega'}&=&\trace(M_B \ketbra{\omega'}).
\end{eqnarray}
This can
trivially be extended to arbitrary $M_A$ \cite{note}. One can recognize now
Werner's model~\cite{W89} where
\begin{eqnarray}
&&\exs{P_k}_{\omega}=\bigg\{
\begin{tabular}{cl}
  $1$ & if $  \langle \omega | P_k | \omega \rangle
  < \langle \omega | P_l | \omega \rangle$ for all $l\ne k$ \\%\nonumber\\
  $0$ & otherwise
\end{tabular}
\label{LHV}
\end{eqnarray}
for party $A$ while
\begin{eqnarray}
\exs{M_B}_{\omega}=\trace(M_B\ketbra{\omega}) \label{LHV2}
\end{eqnarray}
for $B$. This
finishes the proof of Theorem 1.

\section{Three-qubit case}
The previous construction is, in principle, easy to generalize to
more parties as follows
\begin{equation}
    \rho^{(n,c)}:=\int_{\omega \in \mathbb{C}^2,|\omega|=1 }
M(d\omega) \varrho_\omega \otimes \rho_\omega^{\otimes (n-1)} .
\label{decomp3}
\end{equation}
Here, party $A$ receives again a non-physical Bloch vector, while
$B$, $C$,... get a standard qubit. We are now in the position to
prove
the main result of this work.\\
{\it Theorem 2: There exists a LHV model for von Neumann
measurements on states $\rho^{(3,c)}$ for $c\leq 1$. These states
contain genuine three-qubit entanglement if
$c>(\sqrt{13}-1)/3\approx 0.869$.}

{\it Proof of Theorem 2.} Based on our discussion on the two-qubit
case, it is clear that there is a LHV model for von Neumann
measurements on $\rho^{(3,c)}$ if, and this is an important
condition, this state is $U^{\otimes 3}$-invariant. After long but
straightforward calculation, one obtains
\begin{eqnarray}
\rho^{(3,c)}&=&\frac{1}{8} \eins \otimes \eins \otimes \eins +
\sum_{k=x,y,z} \frac{1}{24} \eins \otimes \sigma_k \otimes
\sigma_k \nonumber\\ &-& \frac{c}{16} ( \sigma_k \otimes \eins
\otimes \sigma_k + \sigma_k \otimes \sigma_k \otimes \eins).
\label{rhota2}
\end{eqnarray}
This state is  invariant $U\otimes U\otimes U$ by inspection if we
know that $\sum_k \sigma_k \otimes \sigma_k$ is $U\otimes U$
invariant. Thus correlation measurements on this state fit the LHV
model given in Eqs.~(\ref{LHV}-\ref{LHV2}) for parties $A/B$, while
we have $\exs{M_{C}}_{\omega}=\trace(M_{C}\ketbra{\omega})$ for $C.$

\begin{figure}
%\narrowtext
\centerline{\epsfxsize 2.8in \epsffile{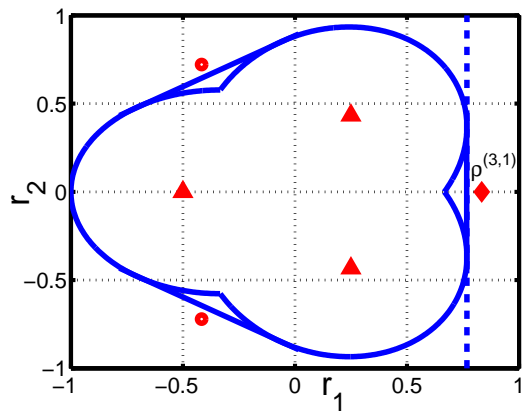} }
\caption{Union of three solid ellipses: Projection of the points
corresponding to pure biseparable states on the $r_1/r_2$ plane.
Together with the solid straight lines: Set of mixed biseparable
states. (diamond) State $\rho^{(3,1)}$. (circles) States obtained
from $\rho^{(3,1)}$ by permuting its qubits. (triangles) Two-qubit
Werner state of the form $\rho^{(2,1)}\otimes\openone/2$
 and states obtained from it by permuting the qubits. For
more details see text.} \label{fig_bisep}
\end{figure}

We now show that $\rho^{(3,c)}$ is genuine three-partite entangled
when $c>(\sqrt{13}-1)/3$, i.e., it cannot be constructed by mixing
pure states with two-qubit entanglement. In particular, it cannot be
constructed by mixing different bipartite Werner states of the form
$\rho^{(2,c)}\otimes\rho_C$, $\rho_A\otimes\rho^{(2,c)}$, etc. In
what follows, we adopt the definitions of  Ref.~\cite{EW01}:
\begin{eqnarray}
R_1&:=&\frac{1}{3}(2V_{BC}-V_{CA}-V_{AB}),\nonumber\\
R_2&:=&\frac{1}{\sqrt{3}}(V_{AB}-V_{CA}),
\end{eqnarray}
where $V_{kl}$ exchanges two qubits, and we use the notation
$r_{k}:=\exs{R_{k}}.$ In order to examine the entanglement
properties of the three-qubit states $\rho^{(3,c)}$, we consider the
projection of the set of
%$U\otimes U\otimes U$ invariant
biseparable states on the $r_1$/$r_2$ plane~\cite{EW01}, as shown in
Fig.~\ref{fig_bisep}. The union of the three solid disks corresponds
to the union of biseparable pure states of the three possible
bipartitioning. One of these disks has the equation \cite{EW01}
\begin{equation}
\bigg(\frac{\sqrt{3}}{2}r_1+\frac{1}{2\sqrt{3}}\bigg)^2+r_2^2 \le
\frac{1}{3}. \label{circle}
\end{equation}
The other two can be obtained through $\pm 120$ degree rotations
around the origin. The straight solid lines indicate the boundary of
the convex hull of these sets. Any biseparable mixed state
corresponds to a point within this set. Based on Eq.~(\ref{circle})
and its rotated versions one can see that for biseparable states
\begin{equation}
\exs{R_1}\le\frac{\sqrt{13}+1}{6}\approx 0.77 . \label{WW}
\end{equation}
In Fig.~\ref{fig_bisep} points fulfilling Eq.~(\ref{WW}) are on the
left hand side of the dashed vertical line. For $\rho^{(3,c)}$ we
have $r_1=(2+3c)/6$ thus the state is genuine three-party entangled
if $c>(\sqrt{13}-1)/3.$ This finishes the proof of Theorem 2.

For the state $\rho^{(3,c)}$ the reduced two-qubit states
$\rho_{AB}$ and $\rho_{AC}$ are entangled if $c>2/3.$ This means
that for $c>2/3$ a singlet between $A$ and $B$, and also between $A$
and $C$ can be distilled. Hence from many copies of our state any
three-qubit state can be obtained with local operations and
classical communication.

\section {States $\rho^{(n,c)}$ and $U^{\otimes n}$-invariant
symmetric extensions of two-qubit Werner states}
At first sight,
there is no reason to limit our discussion to $n=3$. Direct
calculation, however, shows that $\rho^{(4,1)}$ is not $U^{\otimes
4}$ invariant, so the previous local model cannot be applied.
Nevertheless, the states of Eq. (\ref{decomp}) can be used to
construct symmetric extensions \cite{W89Ext,TEPhD}. Recall that a
$(1,n-1)$ symmetric extension of a two-party state $\rho$ is an
$n$-party operator $H_\rho$ such that $\rho=\trace_{3,4,...,n}
H_\rho$ and $H_\rho$ is symmetric under the permutation of parties
$2,3,...,n$. For being an extension, we also need that $H_\rho$ is
positive semidefinite~\cite{W89Ext} and for a quasi-extension that
it is positive on product states \cite{TD02}.

It can be seen by inspection that if $\rho^{(n,c)}$ is positive
semidefinite then it is an extension of $\rho^{(2,c)}.$ Symmetric
extensions can be constructed for a larger range of $c$ if the
matrix $\rho^{(n,c)}$ is twirled:
\begin{eqnarray}
\rho_{T}^{(n,c)}&:=&\int_{U \in U(2)} dU {U^\dagger}^{\otimes
n}\rho^{(n,c)}U^{\otimes n}\nonumber\\&=& \int_{\omega \in
\mathbb{C}^2,|\omega|=1 } M(d\omega) \tau_\omega \otimes \rho_\omega
^{\otimes (n-1)}, \label{decompT}
\end{eqnarray}
where $dU$ denotes the Haar measure, $\rho_\omega$ is defined in
Eq.~(\ref{rhoomega}), and
\begin{equation}
\tau_\omega:=\frac{1}{2}
[\eins-\frac{3}{2}c\sum_{k=x,y,z}\exs{\sigma_k}_\omega\sigma_k].
\label{rhoomegaT}
\end{equation}
Direct calculation shows that for Werner states with $p=2/3$, $5/9$,
and $1/2,$ $(1,m)$ symmetric extensions can be obtained from
Eq.~(\ref{decompT}) for $m=2$, 3, and 4, respectively.

\section {Generalization to more parties or higher dimension}
Now rather than fixing the quantum state from the very beginning, we
will look for the four-qubit quantum state for which the
correlations fit the local model of Eqs.~(\ref{LHV}-\ref{LHV2}),
when party $C$ and $D$ also get physical qubits. Note however that
there is no {\it a priori} reason why a LHV model should give
correlations compatible with a quantum state. The desired state must
be $U^{\otimes4}$ invariant thus it must be a linear combination of
the $4!=24$ permutation operators \cite{EW01}. It must fit all
three-qubit correlations of our LHV model and must be invariant
under the permutation of qubits $B$, $C$ and $D$. It can be proved
that this state must have the form $\rho':=\rho_{N4}- K \{ 3
\sum_{k=x,y,z}\sigma_k \otimes \sigma_k \otimes \sigma_k \otimes
\sigma_k + \sum_{l<k} \Pi[\sigma_k \otimes \sigma_k \otimes \sigma_l
\otimes \sigma_l ] \},$ where $\Pi[A]$ denotes the sum of all
distinct permutations of $A$ and $K$ is a constant. Here $\rho_{N4}$
contains the terms which do not affect four-qubit correlations of
the form $\exs{\sigma_a \otimes \sigma_b\otimes \sigma_c\otimes
\sigma_d}.$ Setting $K=1/128$ our state gives the same four-qubit
correlations as the LHV model for $M_{A/B}=\sigma_x$ and
$M_{C/D}=\sigma_y.$ Due to the finite number of free parameters
there is only one such Hermitian matrix with unit trace. However,
for $M_{A/B/C/D}=\sigma_x$ this matrix fails to reproduce
correlations of the LHV model (i.e., -1/4). Thus, there is not a
matrix corresponding to the $n$-qubit version of the LHV model given
in Eqs.~(\ref{LHV}-\ref{LHV2}) for $n\geq 4$ qubits.

Finally, one can explore whether the local model
Eqs.~(\ref{LHV}-\ref{LHV2}), with
$\exs{M_{C}}_{\omega}=\trace(M_{C}\ketbra{\omega})$ can be
associated to a three-qudit state. Surprisingly, it turns out that
the model obtained this way is not a valid LHV model for a quantum
state when $d>2$. In order to see this, let us consider the case
$d=3.$ Take $M_A=\ketbra{1}$, defined by $\{\alpha_k\}=\{1,0,0\}$
and $\{P_k\}=\{\ketbra{1},\ketbra{2},\ketbra{3}\}$, and
$M'_A=\ketbra{1}$, which is actually equal to $M_A$ but defined by
$\{\alpha'_k\}=\{1,0,0\}$ and
$\{P_k'\}=\{\ketbra{1},\ketbra{2'},\ketbra{3'}\}$, where
$\ket{2'}=\alpha\ket{2}+\beta\ket{3}$ and
$\ket{3'}=\beta^*\ket{2}-\alpha^*\ket{3}$, with
$|\alpha|^2+|\beta|^2=1.$ Moreover, on the other two qudits we
measure $M_B=M_C=\ketbra{2}.$ Using the methods of Ref.~\cite{W89},
we obtain $\exs{M_A \otimes M_B \otimes M_C}=13/162$, while
$\exs{M_A' \otimes M_B \otimes M_C}=15/162$ for
$\alpha=\beta=1/\sqrt{2}.$ Thus
%\begin{equation}
    $\exs{M_A \otimes M_B \otimes M_C}\neq \exs{M'_A \otimes M_B
\otimes
    M_C}.$ %\label{threecorr}
%\end{equation}
A similar lack of selfconsistency can be found for $d>3.$

\section{Conclusions}
We presented a family of three-qubit states for which correlations
for all von Neumann measurements can be described by a LHV model. We
proved that some of these states are genuine three-qubit entangled
and distillable, so three-qubit entanglement is not sufficient for a
state to be nonlocal. We also showed that there is not a quantum
state corresponding to our model with more parties or higher
dimension. For the details of our calculation, see the Appendix of
Ref.~\cite{quantph}. In the future, it would be interesting to extend
our model to general measurements.

\acknowledgments

We thank F. M. Spedalieri and M. M. Wolf for many useful
discussions. We acknowledge the support of the EU projects RESQ and
QUPRODIS and the Kom\-pe\-tenz\-netz\-werk
Quan\-ten\-in\-for\-ma\-ti\-ons\-ver\-ar\-beitung der
Bay\-e\-ri\-schen Staats\-re\-gie\-rung. G.T. thanks the support of
the European Union (Grants Nos. MEIF-CT-2003-500183 and No.
MERG-CT-2005-029146) and the National Research Fund of Hungary  OTKA
under Contract No. T049234. A.A.  thanks the MPQ, Garching, for
hospitality, and acknowledges the Spanish ``Ram\'on y Cajal" grant
(MEC).

% Put in comment if Appendix is needed
%\end{document}

\newpage
\clearpage
\newpage

\section*{Appendix}
In the Appendix we present some details of our computations.
First we show how to compute the integral is
Eq.~(\ref{decomp3}). It can easily be
integrated numerically. For computing the integral
analytically, we can rewrite it as an integration
on the Bloch sphere
\begin{eqnarray}
\rho^{(n,c)}=\int_{\Omega \in \mathbb{R}^3, |\Omega|=1}
M(d\Omega) \varrho_\Omega \otimes \rho_\Omega^{(n-1)},
\label{decompOmega}
\end{eqnarray}
where
\begin{eqnarray}
\varrho_\Omega&:=&\frac{1}{2}(\eins-c{\rm sign}(\Omega)\vec{\sigma}),\nonumber\\
\rho_\Omega&:=&\frac{1}{2}(\eins+\Omega\vec{\sigma}).\label{decompOmega2}
\end{eqnarray}
Here $\Omega$ is the Bloch vector and ${\rm sign}(\Omega)=[{\rm
sign}(\Omega_x),{\rm sign}(\Omega_y),{\rm sign}(\Omega_z)].$
For computing the integral in Eq.~(\ref{decompOmega}), we can use
the following useful expressions
\begin{eqnarray}
&&\int_{\Omega \in \mathbb{R}^3, |\Omega|=1} M(d\Omega)
(\Omega\vec{\sigma}) ^{\otimes m}
\nonumber\\
&&\;\;\;\;\;\;\;\;= \left\{ \begin{tabular}{cl}
  $0$ & if $m=1$ \\
  $\frac{1}{3}\sum_{k=x,y,z} \sigma_k \otimes \sigma_k $ & if $m=2$ \\
  $0$ & if $m=3$,
\end{tabular}\right.\nonumber\\
&&\int_{\Omega \in \mathbb{R}^3, |\Omega|=1} M(d\Omega)
(\Omega\vec{\sigma}) ^{\otimes m} \otimes {\rm sign}(\Omega)
\nonumber\\
&&\;\;\;\;\;\;\;\;= \left\{
\begin{tabular}{cl}
  $0$ & if $m=0$ \\
  $\frac{1}{2}\sum_{k=x,y,z} \sigma_k \otimes \sigma_k$ & if $m=1$ \\
  $0$ & if $m=2$ \\
  $\frac{1}{4}\sum_{k=x,y,z} \sigma_k ^{\otimes 4}+\frac{1}{8} J$ & if
  $m=3,$
\end{tabular}\right.\nonumber\\
&&J=\sum_{l<k} \Pi [\sigma_k \otimes \sigma_k \otimes \sigma_l
\otimes \sigma_l ].
\label{expressions}
\end{eqnarray}
Here $\Pi[A]$ denotes the sum of all different operators obtained from $A$
after permuting the qubits. Based on these, the state corresponding to Eq.~(\ref{decomp3}) for
$n=3$ is Eq.~(\ref{rhota2}). For $n=4$ the density matrix obtained from
Eq.~(\ref{decomp3}) is
\begin{eqnarray}
\rho^{(4,c)}&=&\frac{1}{16} \eins \otimes \eins \otimes \eins
\otimes \eins \nonumber\\ &-& \frac{c}{32} \sum_{k=x,y,z}
\sigma_k\otimes
\Pi[\sigma_k \otimes \eins\otimes \eins] \nonumber\\
 &+&
\frac{1}{48} \eins \otimes
\Pi [ \sigma_k \otimes \sigma_k \otimes \eins ]
 \nonumber\\
&-& \frac{c}{64} \sigma_k \otimes \sigma_k \otimes \sigma_k
\otimes \sigma_k \nonumber\\
&-&
 \frac{c}{128} \sum_{l<k} \Pi[\sigma_k \otimes \sigma_k
\otimes \sigma_l  \otimes \sigma_l].
\end{eqnarray}
This matrix is not $U\otimes U\otimes U\otimes U$ invariant.
However, if for party $A$ the operators $M_A=\sigma_{x/y/z}$ are
measured then the many-body correlations still fit the LHV model
Eqs.~(\ref{LHV},\ref{LHV2}) when for parties $C/D$ we have
$\exs{M_{C/D}}_\omega=Tr(M_{C/D}\ketbra{\omega}).$ For
$\rho^{(4,1)}$ we have
$\exs{\sigma_x\otimes\sigma_x\otimes\sigma_y\otimes\sigma_y}=-1/8$
and
$\exs{\sigma_x\otimes\sigma_x\otimes\sigma_x\otimes\sigma_x}=-1/4.$
After twirling, $\rho^{(4,c)}$ becomes
\begin{eqnarray}
\rho^{(4,c)}_T&=&\frac{1}{16} \eins \otimes \eins \otimes \eins
\otimes \eins \nonumber\\ &-& \frac{c}{32} \sum_{k=x,y,z}
\sigma_k\otimes
\Pi[\sigma_k \otimes \eins\otimes \eins] \nonumber\\
 &+&
\frac{1}{48} \eins \otimes \Pi [ \sigma_k \otimes \sigma_k \otimes
\eins ]
 \nonumber\\
&-& \frac{3c}{160} \sigma_k \otimes \sigma_k \otimes \sigma_k
\otimes \sigma_k \nonumber\\
&-&
 \frac{c}{160} \sum_{l<k} \Pi[\sigma_k \otimes \sigma_k
\otimes \sigma_l  \otimes \sigma_l].
\end{eqnarray}

In the second part of this Appendix we show how the
correlations for the three-qutrit case  were computed. Let us
use the notation $\ket{\omega}=\sum_k
\sqrt{u_k}\exp(i\phi_k)\ket{k'},$ where $u_k$ are non-negative,
$\phi_k$ are real and $\ket{1'}=\ket{1}.$ Hence
$u_k=\exs{\omega|P_k'| \omega}.$ Based on Ref.~\cite{W89} we can
write
\begin{eqnarray}
\exs{M_A' \otimes M_B \otimes M_C}&=& \int_S M(d\omega) \langle \omega| P_2 | \omega \rangle^2,
\label{mabcint}
\end{eqnarray}
where $S$ denotes the subset of $\{\omega \in \mathbb{C}^2,
|\omega|=1\}$ for which $u_1<u_2$ and $u_1<u_3.$ Now, $P_2$ in the
$\{\ket{2'},\ket{3'}\}$ basis is
\begin{equation}
P_2=\bigg[\begin{array}{cc}
|\alpha|^2 & \alpha^* \beta^* \\
\alpha\beta & |\beta|^2
\end{array}\bigg].\label{P2}
\end{equation}
Substituting Eq.~(\ref{P2}) into Eq.~(\ref{mabcint}) we obtain
\begin{eqnarray}
&&\exs{M_A' \otimes M_B \otimes M_C}= \int_S M(d\omega) \bigg[|\alpha|^2 u_2 + |\beta|^2 u_3\nonumber\\
&&\;\;\;\;\;\;\;\;\;\;\;\;+2|\alpha\beta|\sqrt{u_2u_3}\cos(\phi_2-\phi_3+\phi_\alpha+\phi_\beta)\bigg]^2,\nonumber\\
\end{eqnarray}
where $\phi_\alpha$ and $\phi_\beta$ are the phase of $\alpha$ and $\beta,$ respectively.
Now we transform $\int_S M(d\omega)$ into an integral over
$u_k$ and $\phi_k$ and do the integration \cite{W89}. Thus we obtain
\begin{eqnarray}
\exs{M_A' \otimes M_B \otimes M_C}
&=&\frac{13}{162}(|\alpha|^4+|\beta|^4)+\frac{17}{81}|\alpha\beta|^2,\nonumber\\
\end{eqnarray}
while
\begin{eqnarray}
\exs{M_A \otimes M_B \otimes M_C}
&=&\frac{13}{162}.
\end{eqnarray}
Clearly $\exs{M_A' \otimes M_B \otimes M_C}\ne \exs{M_A \otimes M_B \otimes M_C}$ if $0<|\alpha|<1.$

\vfill

\eject


\begin{thebibliography}{99}
\bibitem{note}
In particular, for $M_A=c_1\openone+c_2\hat{n}_A\vec{\sigma}$ we
have $\exs{M_A}_{\omega'}=c_1-c_2{\rm sign}[{\rm
Tr}(\hat{n}_A\vec{\sigma}\ketbra{\omega'})].$
\end{thebibliography}
\end{document}